\documentclass[showpacs,preprintnumbers,amsmath,amssymb]{revtex4}
\usepackage{graphicx}
\usepackage{dcolumn}
\usepackage{bm}
\usepackage{amssymb}
\usepackage{array}
\usepackage{hhline}
\usepackage{longtable}

\begin{document}

\title{Recursive solutions for Laplacian spectra and eigenvectors of\\ a class of growing treelike networks}

\author{Zhongzhi Zhang$^{1,2}$}
\email{zhangzz@fudan.edu.cn}

\author{Yi Qi$^{1,2}$}

\author{Shuigeng Zhou$^{1,2}$}
\email{sgzhou@fudan.edu.cn}

\author{Yuan Lin$^{1,2}$}

\author{Jihong Guan$^{3}$}
\email{jhguan@tongji.edu.cn}

\affiliation {$^{1}$School of Computer Science, Fudan University,
Shanghai 200433, China}

\affiliation {$^{2}$Shanghai Key Lab of Intelligent Information
Processing, Fudan University, Shanghai 200433, China}

\affiliation{$^{3}$Department of Computer Science and Technology,
Tongji University, 4800 Cao'an Road, Shanghai 201804, China}

\begin{abstract}

The complete knowledge of Laplacian eigenvalues and eigenvectors of
complex networks plays an outstanding role in understanding various
dynamical processes running on them; however, determining
analytically Laplacian eigenvalues and eigenvectors is a theoretical
challenge. In this paper, we study the Laplacian spectra and their
corresponding eigenvectors of a class of deterministically growing
treelike networks. The two interesting quantities are determined
through the recurrence relations derived from the structure of the
networks. Beginning from the rigorous relations one can obtain the
complete eigenvalues and eigenvectors for the networks of arbitrary
size. The analytical method opens the way to analytically compute
the eigenvalues and eigenvectors of some other deterministic
networks, making it possible to accurately calculate their spectral
characteristics.

\end{abstract}

\pacs{02.10.Yn, 89.75.Hc, 02.10.Ud}

\date{\today}
\maketitle

\section{Introduction}

As an interdisciplinary subject, complex networks have received
tremendous recent interest from the scientific
community~\cite{AlBa02,DoMe02,Ne03,BoLaMoChHw06,DoGoMe08}, because
of their flexibility and generality in the description of natural
and manmade systems. A central issue in the study of complex
networks is to understand how their dynamical behaviors are
influenced by the underlying topological
structure~\cite{Ne03,BoLaMoChHw06,DoGoMe08}. In various dynamical
processes, the effect of network structure is encoded in the
eigenvalues (spectra) and their corresponding eigenvectors of its
Laplacian matrix. For instance, the synchronizability of a network
is determined by the ratio of the maximum eigenvalue to the smallest
nonzero one of its Laplacian matrix~\cite{BaPe02,ArDiKuMoZh08}.
Again, for example, for continuous-time quantum walks~\cite{FaGu98}
in a network, the quantum transition probabilities~\cite{MuVoBl05}
between two nodes are closely related to the eigenvalues and
orthonormalized eigenvectors of its Laplacian matrix, which also
determine the resistance between a pair of nodes and the average
resistance of all couples of nodes in a resistance
network~\cite{Wu04,KoHaBaBeKoAb06}. Thus, the complete (exact)
knowledge of Laplacian spectra and eigenvectors is very important
for understanding the network dynamics.

Recently, a lot of activities have been devoted to the study of the
spectra of complex
networks~\cite{FaDeBaVi01,GoKaKi01,DoGoMeSa03,ChLuVu03}, providing
useful insight into the topological properties of and dynamical
processes on networks. However, most previous related studies have
been confined to approximate or numerical methods, the latter of
which is prohibitively difficult for large networks because of the
limit of time and memory. Moreover, notwithstanding its
significance, relevant research on eigenvectors of Laplacian matrix
of complex networks is much less.

In the present paper, we investigate the Laplacian eigenvalues and
eigenvectors of a class of deterministic treelike networks, which
are constructed iteratively~\cite{JuKiKa02}. By applying the
technique of graph theory and an algebraic iterative procedure, we
derive recursive relations for the Laplacian eigenvalues and
eigenvectors of the networks. The obtained recurrence relations
allow one to determine explicitly the full Laplacian eigenvalues and
eigenvectors of the considered networks of arbitrary iterations from
those of its initial structure.

\section{Model for the growing trees}

Here we introduce a model for a class of deterministically growing
trees (networks) defined in an iterative way~\cite{JuKiKa02}, which
has attracted an amount of attention~\cite{DoMeOl06,DoGoMe08}. We
investigate this model because of its intrinsic interest and its
deterministic construction, which allows one to study analytically
its Laplacian spectra and their corresponding eigenvectors.

\begin{figure}
\begin{center}
\includegraphics[width=0.85\linewidth,trim=60 05 60 0]{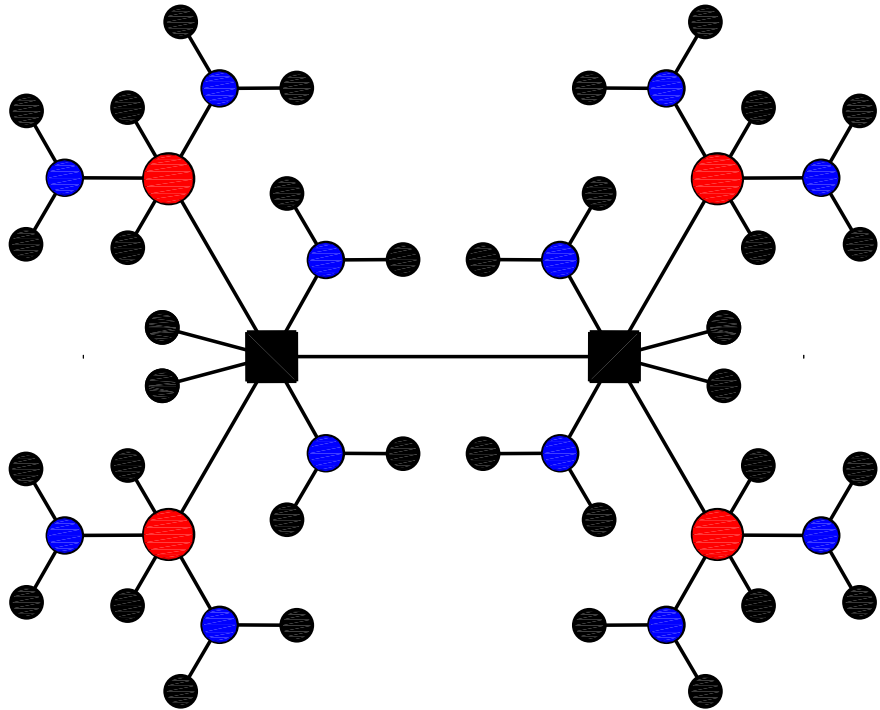}
\caption{(Color online) Illustration of a deterministic uniform
recursive tree for the special case of $m=2$, showing the first
several steps of growth process.} \label{network}
\end{center}
\end{figure}

The deterministically growing trees, denoted by $U_{t}$ ($t\geq 0$)
after $t$ iterations, are constructed as follows. For $t=0$, $U_{0}$
is an edge connecting two nodes. For $t\geq 1$, $U_{t}$ is obtained
from $U_{t-1}$ by attaching $m$ ($m$ is a positive integer) new
nodes to each node in $U_{t-1}$. Figure~\ref{network} illustrates
the construction process of a particular network for the case of
$m=2$ for the first four generations.

According to the network construction, one can see that at each step
$t_i$ ($t_i\geq 1$) the number of newly introduced nodes is $L(t_i)
=2\,m\,(m +1)^{ t_i -1}$. From this result, we can easily compute
the network order (i.e., the total number of nodes) $N_t$ at step
$t$,
\begin{equation}\label{Nt}
N_t=\sum_{t_i=0}^{t}L(t_i)= 2\,(m +1)^{t}\,.
\end{equation}

The considered networks have a degree distribution of exponential
form. Their cumulative degree distribution $P_{\rm cum}(k)$, defined
to be the probability that the degree is greater than or equal to
$k$, decays exponentially with $k$ as $P_{\rm cum}(k)
=(m+1)^{-\frac{k-1}{m}}$~\cite{JuKiKa02}. Their average path length
(APL), defined as the mean of shortest distance between all pairs of
nodes, increases logarithmically with network order~\cite{DoMeOl06}.
Thus, the networks exhibit small-world behavior~\cite{WaSt98}.

Notice that the particular case of $m=1$ is in fact a deterministic
version of the uniform recursive tree (URT)~\cite{SmMa95}, which is
a principal model~\cite{DoKrMeSa08,ZhZhZhGu08} of random
graphs~\cite{ErRe60}.
As one of the most widely studied models, the URT is constructed as
follows~\cite{SmMa95}: start with a single node, at each time step,
we attach a new node to an existing node selected at random. It has
found many important applications in various areas. For example, it
has been suggested as models for the spread of
epidemics~\cite{Mo74}, the family trees of preserved copies of
ancient or medieval texts~\cite{NaHe82}, chain letter and pyramid
schemes~\cite{Ga77}, to name but a few. The $m=1$ case of the
networks studied here has similar structural properties as the URT,
thus, we call the considered networks expanded deterministic uniform
recursive trees (EDURTs), which could shed light in better
understanding the nature of the URT.

After introducing the EDURTs, in what follows we will study the
eigenvalues and their corresponding eigenvectors of the Laplacian
matrices of the EDURTs.

\section{Laplacian spectra and their corresponding eigenvectors}

Generally, for an arbitrary graph, it is difficult to determine all
eigenvalues and eigenvectors of its Laplacian matrix, but below we
will show that for $U_t$ one can settle this problem.

\subsection{Eigenvalues}

As known in Eq.~(\ref{Nt}), there are $2(m+1)^t$ vertices in $U_t$.
We denote by $V_t$ the vertex set of $U_t$, i.e.,
$V_t=\{v_1,v_2,\ldots, v_{2(m+1)^t}\}$. Let $\mathbf{A}_t=[a_{ij}]$
be the adjacency matrix of network $U_t$, where $a_{ij} =a_{ji}=1$
if nodes $i$ and $j$ are connected, $a_{ij} =a_{ji}=0$ otherwise,
then the degree of vertex $v_i$ is defined as  $d_{v_i}= \sum_{j\in
V_t}a_{ij} $. Let $\mathbf{D}_t ={\rm diag} (d_{v_1},
d_{v_2},\ldots, d_{v_{{2(m+1)^t}}})$ represent the diagonal degree
matrix of $U_t$, then the Laplacian matrix of $U_t$ is defined by
$\mathbf{L}_t=\mathbf{D}_t-\mathbf{A}_t$.

We first study the Laplacian spectra of $U_t$, while we leave the
eigenvectors to the next subsection. By construction, it is easy to
find that the adjacency matrix $\textbf{A}_t$ and diagonal degree
matrix $\textbf{D}_t$ satisfy the following relations:
\begin{equation}\label{matrix01}
\mathbf{A}_t=\left(\begin{array}{ccccc}\textbf{A}_{t-1} &
\textbf{I}& \textbf{I} &\cdots& \textbf{I}
\\\textbf{I} &{\textbf{0}}& {\textbf{0}}&\cdots& {\textbf{0}}
\\\textbf{I} & {\textbf{0}}& {\textbf{0}}&\cdots& {\textbf{0}}
\\\vdots & \vdots & \vdots & \  &\vdots
\\\textbf{I} & {\textbf{0}}& {\textbf{0}}&\cdots& {\textbf{0}}
\end{array}\right)_{(m+1)\times(m+1)}
\end{equation}
and
\begin{equation}\label{matrix02}
\mathbf{D}_t=\left(\begin{array}{ccccc}\textbf{D}_{t-1}+ m\textbf{I}
& \textbf{0}& \textbf{0} &\cdots& \textbf{0}
\\\textbf{0} &{\textbf{I}}& {\textbf{0}}&\cdots& {\textbf{0}}
\\\textbf{0} & {\textbf{0}}& {\textbf{I}}&\cdots& {\textbf{0}}
\\\vdots & \vdots & \vdots & \  &\vdots
\\\textbf{0} & {\textbf{0}}& {\textbf{0}}&\cdots& {\textbf{I}}
\end{array}\right)_{(m+1)\times(m+1)}\,,
\end{equation}
where each block is a $2(m+1)^{t-1}\times 2(m+1)^{t-1}$ matrix and
$\textbf{I}$ is the identity matrix. Thus, according to the above
expressions (for $\mathbf{A}_t$ and $\mathbf{D}_t$) and the
definition of Laplacian matrix, we have the following recursive
relation between $\mathbf{L}_t$ and $\mathbf{L}_{t-1}$:
\begin{eqnarray}\label{matrix03}
\mathbf{L}_t &=& \mathbf{D}_t-\mathbf{A}_t \nonumber\\
&=&\left(\begin{array}{ccccc}\textbf{L}_{t-1}+m\textbf{I} &
\textbf{-I}& \textbf{-I} &\cdots& \textbf{-I}
\\\textbf{-I} &{\textbf{I}}& {\textbf{0}}&\cdots& {\textbf{0}}
\\\textbf{-I} & {\textbf{0}}& {\textbf{I}}&\cdots& {\textbf{0}}
\\\vdots & \vdots & \vdots & \  &\vdots
\\\textbf{-I} & {\textbf{0}}& {\textbf{0}}&\cdots& {\textbf{I}}
\end{array}\right).
\end{eqnarray}
Then, the characteristic polynomial of $\textbf{L}_t$ is
\begin{eqnarray}\label{matrix04}
P_t(x)&=&{\rm det}(x\textbf{I}-\textbf{L}_t)\nonumber\\
&=&{\rm
det}\left(\begin{array}{ccccc}(x-m)\textbf{I}-\textbf{L}_{t-1} &
\textbf{I}& \textbf{I} &\cdots& \textbf{I}
\\\textbf{I} & (x-1)\textbf{I} & {\textbf{0}}&\cdots& {\textbf{0}}
\\\textbf{I} & {\textbf{0}}& (x-1)\textbf{I} &\cdots& {\textbf{0}}
\\\vdots & \vdots & \vdots & \  &\vdots
\\\textbf{I} & {\textbf{0}}& {\textbf{0}}&\cdots& (x-1)\textbf{I}
\end{array}\right)\nonumber
\\&=&({\rm det}((x-1)\textbf{I}))^m\cdot{\rm det}\left(\begin{array}{ccccc}(x-m)\textbf{I}-\textbf{L}_{t-1}
& \textbf{I}& \textbf{I} &\cdots& \textbf{I}
\\\frac{1}{x-1}\textbf{I} & \textbf{I} & {\textbf{0}}&\cdots& {\textbf{0}}
\\\frac{1}{x-1}\textbf{I} & {\textbf{0}}& \textbf{I} &\cdots& {\textbf{0}}
\\\vdots & \vdots & \vdots & \  &\vdots
\\\frac{1}{x-1}\textbf{I} & {\textbf{0}}& {\textbf{0}}&\cdots& \textbf{I}
\end{array}\right)\nonumber
\\&=&({\rm det}((x-1)\textbf{I}))^m\cdot{\rm det}\left(\begin{array}{ccccc}(x-m-\frac{m}{x-1})\textbf{I}-\textbf{L}_{t-1}
& \textbf{0}& \textbf{0} &\cdots& \textbf{0}
\\\frac{1}{x-1}\textbf{I} & \textbf{I} & {\textbf{0}}&\cdots& {\textbf{0}}
\\\frac{1}{x-1}\textbf{I} & {\textbf{0}}& \textbf{I} &\cdots& {\textbf{0}}
\\\vdots & \vdots & \vdots & \  &\vdots
\\\frac{1}{x-1}\textbf{I} & {\textbf{0}}& {\textbf{0}}&\cdots& \textbf{I}
\end{array}\right)\nonumber,\nonumber\\
\end{eqnarray}
where the elementary operations of matrix have been used. According
to the results in~\cite{Si00}, we have
\begin{eqnarray}\label{matrix05}
P_t(x)=\big({\rm det}((x-1)\textbf{I})\big)^m \times {\rm
det}\left(\left(x-m-\frac{m}{x-1}\right)\textbf{I}-\textbf{L}_{t-1}\right).
\end{eqnarray}
Thus, $P_t(x)$ can be recast recursively as follows:
\begin{equation}\label{matrix06}
P_t(x)=(x-1)^{2m(m+1)^{t-1}}\times P_{t-1}(\varphi(x)),
\end{equation}
where $\varphi(x)=x-m-\frac{m}{x-1}$. This recursive relation given
by Eq.~(\ref{matrix06}) is very important, from which we will
determine the complete Laplacian eigenvalues of $U_t$ and their
corresponding eigenvectors. Notice that $P_{t-1}(x)$ is a monic
polynomial of degree $2(m+1)^{t-1}$, then the exponent of
$\frac{m}{x-1}$ in $P_{t-1}(\varphi(x))$ is $2(m+1)^{t-1}$, and
hence the exponent of factor $x-1$ in $P_t(x)$ is
\begin{equation}
2m(m+1)^{t-1}-2(m+1)^{t-1}=2(m-1)(m+1)^{t-1}.
\end{equation}
Consequently, $1$ is an eigenvalue of $\textbf{L}_t$, and its
multiplicity is $2(m-1)(m+1)^{t-1}$.

Note that $U_t$ has $2(m+1)^t$ Laplacian eigenvalues. We represent
these $2(m+1)^t$ Laplacian eigenvalues as $\lambda^t_1,
\lambda^t_2,\dots,\lambda^t_{2(m+1)^t}$, respectively. For
convenience, we presume $\lambda^t_1 \le \lambda^t_2 \le \dots \le
\lambda^t_{2(m+1)^t}$, and denote by $E_t$ the set of these
Laplacian eigenvalues, i.e. $E_t=\{\lambda^t_1,
\lambda^t_2,\dots,\lambda^t_{2(m+1)^t}\}$. All the Laplacian
eigenvalues in set $E_t$ can be divided into two parts. According to
the above analysis, $\lambda=1$ is a Laplacian eigenvalue with
multiplicity $2(m-1)(m+1)^{t-1}$, which gives a part of the
eigenvalues of $L_t$. We denote by $E^{'}_t$ the set of Laplacian
eigenvalues 1 of $U_t$, i.e.,
\begin{equation}
E^{'}_t=\{\underbrace{1,1,1,\dots,1,1}_{2(m-1)(m+1)^{t-1}\mbox{}}\}
\end{equation}
It should be noted that here we neglect the distinctness of elements
in the set. The remaining $4(m+1)^{t-1}$ Laplacian eigenvalues of
$U_t$ are determined by the equation $P_{t-1}(\varphi(x))=0$. Let
the $4(m+1)^{t-1}$ eigenvalues be $\tilde{\lambda}^t_1,
\tilde{\lambda}^t_2,\dots,\tilde{\lambda}^t_{4(m+1)^{t-1}}$,
respectively. For convenience, we presume $\tilde{\lambda}^t_1 \le
\tilde{\lambda}^t_2 \le \dots \le \tilde{\lambda}^t_{4(m+1)^{t-1}}$,
and denote by $E^{*}_t$ the set of these eigenvalues, i.e.,
$E^{*}_t=\{\tilde{\lambda}^t_1,
\tilde{\lambda}^t_2,\dots,\tilde{\lambda}^t_{4(m+1)^{t-1}}\}$.
Therefore, the set of all Laplacian eigenvalues for $U_t$ can be
expressed as $E_t=E^{'}_t \cup E^{*}_t$.

According to Eq.~(\ref{matrix06}), for an arbitrary element in
$E_{t-1}$, say $\lambda_{i}^{t-1} \in E_{t-1}$, both solutions of
$x-m-\frac{m}{x-1}=\lambda_{i}^{t-1}$ are in $E^{*}_t$. In fact,
equation $x-m-\frac{m}{x-1}=\lambda_{i}^{t-1}$ is equivalent to
\begin{equation}\label{matrix07}
x^2-(\lambda_{i}^{t-1}+m+1)x+\lambda_{i}^{t-1}=0.
\end{equation}
We use notations $\tilde{\lambda}_{i}^{t}$ and
$\tilde{\lambda}_{i+2(m+1)^{t-1}}^{t}$ to represent the two
solutions of Eq.~(\ref{matrix07}), since they provide a natural
increasing order of the eigenvalues of $U_t$, which can be seen from
the argument below. Solving this quadratic equation, its roots are
obtained to be $\tilde{\lambda}_{i}^{t}=r_1(\lambda_{i}^{t-1})$ and
$\tilde{\lambda}_{i+2(m+1)^{t-1}}^{t}=r_2(\lambda_{i}^{t-1})$, where
the functions $r_1(\lambda)$ and $r_2(\lambda)$ satisfy
\begin{eqnarray}
r_1(\lambda)=\frac{1}{2}\left(\lambda+m+1-\sqrt{(\lambda+m+1)^2-4\lambda}\right),\label{matrix08}\\
r_2(\lambda)=\frac{1}{2}\left(\lambda+m+1+\sqrt{(\lambda+m+1)^2-4\lambda}\right).\label{matrix00}
\end{eqnarray}
Substituting each Laplacian eigenvalue of $U_{t-1}$ into Eqs.
(\ref{matrix08}) and~(\ref{matrix00}), we can obtain the subset
$E^{*}_{t}$ of Laplacian eigenvalues of $U_{t}$. Since
$E_{0}=\{0,2\}$, by recursively applying the functions provided by
Eqs. (\ref{matrix08}) and~(\ref{matrix00}), the Laplacian spectra of
$U_{t}$ can be determined completely.

It is obvious that both $r_1(\lambda)$ and $r_2(\lambda)$ are
monotonously increasing functions, and that they lie in intervals
$[0,1)$ and $(1,+\infty)$, respectively. On the other hand, since
$r_1(\lambda)-1=\frac{1}{2}\left(\lambda+m-1-\sqrt{(\lambda+m-1)^2+4m}\right)<0$,
we have $r_1(\lambda)<1$. Similarly, we can show that
$r_2(\lambda)>1$. Thus for arbitrary fixed $\lambda'$, $r_1(\lambda)
<1< r_2(\lambda')$ holds for all $\lambda$. Then we have the
following conclusion: If the set of Laplacian eigenvalues for
$U_{t-1}$ is
$E_{t-1}=\{\lambda^{t-1}_1,\lambda^{t-1}_2,\dots,\lambda^{t-1}_{2(m+1)^{t-1}}\}$,
then solving Eqs.~(\ref{matrix08}) and (\ref{matrix00}) one can
obtain the subset $E^{*}_t$ of Laplacian eigenvalues for $U_t$ to be
$E^{*}_t=\{\tilde{\lambda}^t_1,
\tilde{\lambda}^t_2,\dots,\tilde{\lambda}^t_{4(m+1)^{t-1}}\}$, \
where $\tilde{\lambda}^t_1 \le \tilde{\lambda}^t_2 \le \dots \le
\tilde{\lambda}^t_{2(m+1)^{t-1}} < 1 <
\tilde{\lambda}^t_{2(m+1)^{t-1}+1} \le
\tilde{\lambda}^t_{2(m+1)^{t-1}+2} \le \dots \le
\tilde{\lambda}^t_{4(m+1)^{t-1}}$. Recall that $E^{'}_t$ consists of
$2(m-1)(m+1)^{t-1}$ elements, all of which are $1$, so we can easily
get the set of eigenvalue spectra for $U_t$ to be $E_t=E^{*}_t \cup
E^{'}_t$.

From above arguments, it is easy to see that for the special case of
$m=1$, all the $2^{t+1}$ Laplacian eigenvalues of $U_t$ are
fundamentally distinct, which is an interesting property and has
never (to the best of our knowledge) been previously reported in
other network models thus may have some far-reaching consequences.
For other cases $m>1$, some eigenvalues (e.g., 1) are multiple,
which is obviously different from that of $m=1$ case.

It has been established that Laplacian eigenvalues have connections
with many contexts in the theory of networks. For example, they are
closely related to the number of spanning trees on complex
networks~\cite{Bo98}. It has been shown that the number of spanning
tress on a connected network $G$ with order $N$, $N_{\rm{st}}(G)$,
concerns with all its nonzero Laplacian eigenvalues $\lambda_i$
(assuming $\lambda_1=0$ and $\lambda_i\neq 0$ for $i=2,\cdots, N$),
obeying the following expression~\cite{TzWu00}
\begin{equation}\label{spanning}
N_{\rm{st}}(G)=\frac{1}{N}\prod_{i=2}^{N}\lambda_i\,.
\end{equation}
Since $U_t$ are trees for all parameter $m$, according to
Eq.~(\ref{spanning}), the product of all nonzero Laplacian
eigenvalues for $U_t$, denoted by $\Lambda_t$, should be equal to
$N_t$, which can be confirmed from the following argument. For
$t=0$, by construction it is obvious that $\Lambda_0=N_0=2$; for
$t\geq 1$, according to Eq.~(\ref{matrix07}), we can easily obtain
the following recursive relation $\Lambda_t=(m+1)\,\Lambda_{t-1}$,
which combining with the initial value $\Lambda_0=2$ leads to
$\Lambda_t=2(m+1)^t=N_t$. This proves that our computation on the
Laplacian eigenvalues for $U_t$ is right.

\subsection{Eigenvectors}

Similar to the eigenvalues, the eigenvectors of $\textbf{L}_t$
follow directly from those of $\textbf{L}_{t-1}$. Assume that
$\lambda$ is an arbitrary Laplacian eigenvalue of $U_t$, whose
corresponding eigenvector is $\textbf{\emph{v}} \in
\textbf{R}^{2(m+1)^t}$, where $\textbf{R}^{2(m+1)^t}$ represents the
$2(m+1)^t$-dimensional vector space. Then we can solve equation
($\lambda\, \textbf{I} -\textbf{L}_t)\textbf{\emph{v}}=0$ to find
the eigenvector $\textbf{\emph{v}}$. We distinguish two cases:
$\lambda \in E^{*}_t$ and $\lambda \in E^{'}_t$, which will be
separately addressed in detail as follows.

For the first case $\lambda \in E^{*}_t$, we can rewrite the
equation ($\lambda\, \textbf{I} -\textbf{L}_t)\textbf{\emph{v}}=0$
as
\begin{equation}\label{vector01}
\left(\begin{array}{ccccc}(\lambda-m)\textbf{I}-\textbf{L}_{t-1} &
\textbf{I}& \textbf{I} &\cdots& \textbf{I}
\\\textbf{I} &(\lambda-1)\textbf{I}& {\textbf{0}}&\cdots& {\textbf{0}}
\\\textbf{I} & {\textbf{0}}& (\lambda-1)\textbf{I}&\cdots& {\textbf{0}}
\\\vdots & \vdots & \vdots & \  &\vdots
\\\textbf{I} & {\textbf{0}}& {\textbf{0}}&\cdots& (\lambda-1)\textbf{I}
\end{array}\right)\left(\begin{array}{c}\textbf{\emph{v}}_1\\\textbf{\emph{v}}_2\\\textbf{\emph{v}}_3 \\\vdots\\ \textbf{\emph{v}}_{m+1}
\end{array}\right)=0,
\end{equation}
where vector $\textbf{\emph{v}}_i$ ($1\le i \le m+1$) are components
of $\textbf{\emph{v}}$. Equation~(\ref{vector01}) results in the
following equations:
\begin{eqnarray}
\big((\lambda-m)\textbf{I}_{t-1}-\textbf{L}_{t-1}\big)\textbf{\emph{v}}_1+\textbf{\emph{v}}_2+\dots+\textbf{\emph{v}}_{m+1}=\textbf{0},\label{vector02}
\\ \textbf{\emph{v}}_1+(\lambda-1)\textbf{\emph{v}}_i=\textbf{0}\ \ \  (2\le i \le m+1).\label{vector03}
\end{eqnarray}
Resolving Eq.~(\ref{vector03}), we find that
\begin{eqnarray}\label{vector04}
\textbf{\emph{v}}_i=-\frac{1}{\lambda-1}\textbf{\emph{v}}_1 \ \
(2\le i \le m+1).
\end{eqnarray}
Substituting Eq.~(\ref{vector04}) into Eq.~(\ref{vector02}) we have
\begin{eqnarray}\label{vector05}
\left[\left(\lambda-m-\frac{m}{\lambda-1}\right)\textbf{I}-\textbf{L}_{t-1}\right]\textbf{\emph{v}}_1=0,
\end{eqnarray}
which indicates that $\textbf{\emph{v}}_1$ is the solution of
Eq.~(\ref{vector02}) while $\textbf{\emph{v}}_i$ ($2\le i\le m+1$)
are uniquely decided by $\textbf{\emph{v}}_1$ via
Eq.~(\ref{vector04}).

In Eq.~(\ref{matrix06}), it is clear that if $\lambda$ is an
eigenvalue of Laplacian matrix $\textbf{L}_t$, then
$f(\lambda)=\lambda-m-\frac{m}{\lambda-1}$ must be one eigenvalue of
$\textbf{L}_{t-1}$. [Recall that if $\lambda=\tilde{\lambda}_i^t \in
E^{*}_t$, then $\varphi(\tilde{\lambda}_i^t)=\lambda_{i}^{t-1}$ for
$i\le 2(m+1)^{t-1}$, or
$\varphi(\tilde{\lambda}_i^t)=\lambda_{i-2(m+1)^{t-1}}$ for
$i>2(m+1)^{t-1}$.] Thus, Eq.~(\ref{vector05}) together with
Eq.~(\ref{matrix06}) shows that $\textbf{\emph{v}}_1$ is an
eigenvector of matrix $\textbf{L}_{t-1}$ corresponding to the
eigenvalue $\lambda-m-\frac{m}{\lambda-1}$ determined by $\lambda$,
while
\begin{equation}
\textbf{\emph{v}}=\left(\begin{array}{ccc}\textbf{\emph{v}}_1
\\\textbf{\emph{v}}_2\\\textbf{\emph{v}}_3\\\vdots\\\textbf{\emph{v}}_{m+1} \end{array}\right)=
\left(\begin{array}{ccc}\textbf{\emph{v}}_1\\
-\frac{1}{\lambda-1}\textbf{\emph{v}}_1\\
-\frac{1}{\lambda-1}\textbf{\emph{v}}_1 \\ \vdots \\
-\frac{1}{\lambda-1}\textbf{\emph{v}}_1\end{array}\right)
\end{equation}
is an eigenvector of $\textbf{L}_t$ corresponding to the eigenvalue
$\lambda$.

Since for the initial graph $U_0$, its Laplacian matrix
$\textbf{L}_0$ has two eigenvalues 0 and 2 with respective
eigenvectors $(1,1)^\top$ and $(1,-1)^\top$; by recursively applying
the above process, we can obtain all the eigenvectors corresponding
to $\lambda \in E^{*}_t$.

For the second case of $\lambda \in E^{'}_t$, where all $\lambda=1$,
the equation ($\lambda\, \textbf{I}
-\textbf{L}_t)\textbf{\emph{v}}=0$ can be recast as
\begin{equation}\label{vector11}
\left(\begin{array}{ccccc}(1-m)\textbf{I}-\textbf{L}_{t-1} &
\textbf{I}& \textbf{I} &\cdots& \textbf{I}
\\\textbf{I} &\textbf{0}& {\textbf{0}}&\cdots& {\textbf{0}}
\\\textbf{I} & {\textbf{0}}& \textbf{0}&\cdots& {\textbf{0}}
\\\vdots & \vdots & \vdots & \  &\vdots
\\\textbf{I} & {\textbf{0}}& {\textbf{0}}&\cdots& \textbf{0}
\end{array}\right)\left(\begin{array}{c}\textbf{\emph{v}}_1\\\textbf{\emph{v}}_2\\\textbf{\emph{v}}_3 \\\vdots\\ \textbf{\emph{v}}_{m+1}
\end{array}\right)=0,
\end{equation}
where vector $\textbf{\emph{v}}_i$ ($1\le i \le m+1$) are components
of $\textbf{\emph{v}}$. Equation~(\ref{vector11}) leads to the
following equations:
\begin{eqnarray}
\textbf{\emph{v}}_1=\textbf{0},\label{vector12}
\\\textbf{\emph{v}}_2+\textbf{\emph{v}}_3+\dots+\textbf{\emph{v}}_{m+1}=\textbf{0}.\label{vector13}
\end{eqnarray}
In Eq.~(\ref{vector12}), $\textbf{\emph{v}}_1$ is a zero vector, and
we denote by  $\textbf{\emph{v}}_{i,j}$ the $j$th component of the
column vector $\textbf{\emph{v}}_i$. On the other hand,
Eq.~(\ref{vector13}) gives us the following equations:
\[ \left \{
\begin{array}{ccccccccc}
\textbf{\emph{v}}_{2,1}&+&\textbf{\emph{v}}_{3,1}&+&\dots
&+&\textbf{\emph{v}}_{m+1,1}=\textbf{0}
\\\textbf{\emph{v}}_{2,2}&+&\textbf{\emph{v}}_{3,2}&+&\dots
&+&\textbf{\emph{v}}_{m+1,2}=\textbf{0}
\\ \vdots & \vdots  &\vdots  & \vdots  &\  &\vdots &\vdots
\\\textbf{\emph{v}}_{2,2(m+1)^{t-1}}&+&\textbf{\emph{v}}_{3,2(m+1)^{t-1}}&+&\dots
&+&\textbf{\emph{v}}_{m+1,2(m+1)^{t-1}}=\textbf{0}
\end{array} \right.
\]

The set of all solutions to any of the above equations consists of
vectors that can be written as
\begin{equation}\label{vector15}
\left(\begin{array}{ccc}\textbf{\emph{v}}_{2,j}
\\\textbf{\emph{v}}_{3,j}\\\textbf{\emph{v}}_{4,j}\\\vdots\\\textbf{\emph{v}}_{m+1,j} \end{array}\right)=
k_{1,j}
\left(\begin{array}{ccc}\textbf{-1}\\
\textbf{1}\\
\textbf{0} \\ \vdots \\
\textbf{0} \end{array}\right)+ k_{2,j}
\left(\begin{array}{ccc}\textbf{-1}\\
\textbf{0}\\
\textbf{1} \\ \vdots \\
\textbf{0} \end{array}\right)+\dots+k_{m-1,j}
\left(\begin{array}{ccc}\textbf{-1}\\
\textbf{0}\\
\textbf{0} \\ \vdots \\
\textbf{1} \end{array}\right),
\end{equation}
where $k_{1,j}$ , $k_{2,j}$ , $\dots$ , $k_{m-1,j}$ are arbitrary
real numbers. In Eq.~(\ref{vector15}), the solutions for all the
vectors $\textbf{\emph{v}}_i$ ($2\le i \le m+1$) can be rewritten as
\begin{equation}\label{vector16}
\left(\begin{array}{c}\textbf{\emph{v}}_2^\top\\\textbf{\emph{v}}_3^\top\\\textbf{\emph{v}}_4^\top
\\\vdots\\ \textbf{\emph{v}}_{m+1}^\top
\end{array}\right)=
\left(\begin{array}{cccc}\textbf{-1} & \textbf{-1}& \cdots&
\textbf{-1}
\\\textbf{1} & {\textbf{0}}&\cdots& {\textbf{0}}
\\\textbf{0} & {\textbf{1}}& \cdots& {\textbf{0}}
\\\vdots & \vdots &  \  &\vdots
\\\textbf{0} & {\textbf{0}}& \cdots& \textbf{1}
\end{array}\right)\\
\\
\left(\begin{array}{cccc}k_{1,1} & k_{1,2}& \cdots&
k_{1,2(m+1)^{t-1}}
\\k_{2,1} &k_{2,2} & \cdots & k_{2,2(m+1)^{t-1}}
\\k_{3,1} & k_{3,2}& \cdots& k_{3,2(m+1)^{t-1}}
\\\vdots  & \vdots & \  &\vdots
\\k_{m-1,1}  & k_{m-1,2} &\cdots& k_{m-1,2(m+1)^{t-1}}
\end{array}\right),
\end{equation}
where $k_{i,j}$ \big($1\le i\le m-1$; $1\le j\le 2(m+1)^{t-1}$\big)
are arbitrary real numbers. According to Eq.~(\ref{vector16}), we
can obtain the eigenvector $\textbf{\emph{v}}$ corresponding to the
eigenvalue $\textbf{1}$. Moreover, it is easy to see that the
dimension of the eigenspace of matrix $\textbf{L}_t$ associated with
eigenvalue $\textbf{1}$ is $2(m-1)(m+1)^{t-1}$.

In this way, all eigenvalues and their corresponding eigenvectors of
$U_t$ have been completely determined in a recursive way.

\section{Conclusions}

In this paper, we have investigated the Laplacian eigenvalues and
their corresponding eigenvectors of a family of deterministically
growing treelike networks that exhibit small-world behavior. Making
use of the methods of linear algebra and graph theory, we have fully
characterized the Laplacian eigenvalues and eigenvectors of the
networks, all of which are recursively determined from those for the
initial network. Interestingly, we showed that for a particular case
($m=1$) of the networks under consideration, all its Laplacian
eigenvalues are disparate. We expect our results to be interesting
in some fields of networks, such as random and quantum walks on
networks, the computation of the resistance between two arbitrary
nodes in a resistor network, the dynamics of coupled oscillators on
networks, and so on.  We also expect that the computing methods of
eigenvalues and eigenvectors used here might be extended to other
types deterministic networks, e.g., deterministic small-world
networks~\cite{CoOzPe00,ZhRoGo06} and deterministic scale-free
networks~\cite{BaRaVi01,DoGoMe02,HiBe06,ZhZhFaGuZh07,RoHaAv07,ZhZhChGu08}.

\subsection*{Acknowledgment}

We would like to thank Yichao Zhang for support. This research was
supported by the National Basic Research Program of China under
grant No. 2007CB310806, the National Natural Science Foundation of
China under Grant Nos. 60704044, 60873040 and 60873070, Shanghai
Leading Academic Discipline Project No. B114, and the Program for
New Century Excellent Talents in University of China (NCET-06-0376).

\end{document}